*The Electoral Consequences of Natural Disasters: A Dynamic Fixed-Effects Analysis*

**Nima Taheri Hosseinkhani**

nt612@rutgers.edu

https://orcid.org/0009-0007-5564-7839

**Finance Department, Rutgers Business School**

**7/16/2025**

**Abstract**

With the increasing frequency of major natural disasters, understanding their political consequences is of paramount importance for democratic accountability. The existing literature is deeply divided, with some studies finding that voters punish incumbents for disaster-related damages, while others find they reward them for relief efforts. This paper investigates the electoral consequences of natural disasters for incumbent mayors, broader electoral dynamics, and the long-term political ambition of officeholders. The study leverages a comprehensive panel dataset of over 10,000 candidate-election observations in U.S. mayoral races from 1989 to 2021, combining detailed election data with a global registry of disaster events. To identify causal effects, the analysis employs a robust dynamic two-way fixed-effects event-study design, validated by extensive pre-trend and placebo tests. The findings reveal that the electoral impact of disasters is highly conditional on their timing. A disaster that strikes in the same quarter as an election provides a significant electoral boost to incumbents, increasing their vote share by over 6 percentage points. However, disasters consistently suppress voter turnout, reducing it by an average of 1.4 percentage points. In a novel finding, the analysis demonstrates that the experience of managing a disaster significantly increases an incumbent's likelihood of seeking re-election in the subsequent cycle by as much as 12 percentage points. These findings help reconcile conflicting theories of retrospective voting by highlighting the critical role of voter myopia and salience. They also reveal a previously undocumented channel through which crises shape political careers, suggesting that disaster management is not only a test of governance but also a catalyst for political ambition.

# Introduction

In recent decades, the United States has experienced a dramatic and costly escalation in the frequency of major weather and climate disasters. The annual average of billion-dollar events has surged from 9.0 in the 1980–2024 period to 23.0 in the last five years alone, with cumulative costs exceeding $2.9 trillion (National Oceanic and Atmospheric Administration, n.d.). This trend makes it a statistical certainty that democratic processes, from voter registration to election day logistics, will be increasingly disrupted and shaped by these exogenous shocks. While the physical and economic impacts of disasters are well-documented, their political consequences remain the subject of a deep and persistent debate in political science, raising fundamental questions about the nature of democratic accountability.

At the heart of this debate is a central puzzle: how do voters respond to events that are, at their origin, outside of any politician's control? One influential school of thought posits that voters engage in "blind retrospection," indiscriminately punishing incumbents for any negative outcome, regardless of fault (Achen & Bartels, 2017). From this perspective, elections are noisy and often irrational affairs where officeholders are held responsible for everything from droughts to shark attacks (Achen & Bartels, 2017). A competing camp argues for a more "attentive electorate" capable of distinguishing between an "act of God" and the government's policy response (Achen & Bartels, 2017; Gasper & Reeves, 2011). These scholars find that voters evaluate the quality and timeliness of relief efforts, rewarding incumbents who demonstrate competence in crisis (Bechtel & Hainmueller, 2011; Gasper & Reeves, 2011; Healy & Malhotra, 2009). This scholarly impasse is not merely academic; it strikes at the core of democratic theory. If voters are blind, elections risk becoming a random lottery where competent leaders are punished for bad luck. If they are attentive, crises can serve as powerful moments of accountability, allowing voters to identify and retain effective leaders.

This paper intervenes in this long-standing debate by arguing that the electoral consequences of natural disasters are more nuanced and dynamic than either of these competing theories suggests. Based on a comprehensive analysis of over three decades of U.S. mayoral elections, this study advances a three-part argument.

First, the electoral fate of incumbents is critically conditional on the *timing* of the disaster. The evidence suggests that voters are neither completely blind nor fully attentive in a comprehensive sense, but are instead highly myopic. Disasters do not lead to a consistent punishment or a reward for competent long-term management. Rather, they trigger a short-lived "rally-'round-the-flag" effect that benefits the incumbent only when the event is highly salient. The analysis shows that a disaster occurring in the immediate run-up to an election—within the same quarter—provides a substantial electoral boost to the incumbent mayor, increasing their vote share by over 6 percentage points. This effect decays rapidly and is statistically nonexistent for disasters occurring earlier in the election cycle.

Second, while recent disasters may help incumbents at the ballot box, they consistently harm the democratic process by depressing citizen participation. The logistical hurdles of voting in a disaster zone—from displacement and damaged infrastructure to the loss of documentation—combined with the immense psychological toll on residents, create significant barriers to participation. The analysis finds that a pre-election disaster reduces voter turnout by a statistically and substantively significant 1.4 percentage points on average.

Third, the impact of a disaster extends beyond the immediate election cycle, reshaping an incumbent's future career calculus. In a novel contribution to the literature on political ambition, this study demonstrates that the experience of managing a crisis serves as a powerful catalyst for future political engagement. Mayors who govern through a major disaster are significantly more likely to run for re-election in the subsequent cycle, with the probability of future participation increasing by as much as 12 percentage points. This suggests that crises are not only tests of leadership but may also reinforce an incumbent's commitment to public service or enhance their perceived political viability.

This paper's contribution is both theoretical and empirical. Theoretically, it offers a more nuanced model of retrospective voting in the context of disasters, one that integrates voter myopia and political ambition to help reconcile the field's conflicting findings. Empirically, it provides some of the most credible causal estimates to date by analyzing a vast, candidate-level panel dataset of U.S. mayoral elections—a critical but understudied level of government where accountability for disaster response is most direct and visible (e.g., Arceneaux & Stein, 2006). The study's dynamic two-way fixed-effects (TWFE) event-study design, buttressed by an extensive series of placebo and pre-trend tests, rigorously addresses the endogeneity and measurement error challenges that have complicated prior research in this area (Bertrand et al., 2004; Imai & Kim, 2021).

The remainder of this paper proceeds as follows. Section 2 reviews the relevant literature on retrospective voting, voter myopia, and political ambition to develop our theoretical framework and hypotheses. Section 3 details the data and empirical strategy. Section 4 presents the main results, and Section 5 discusses the implications of our findings and concludes.

**Literature Review: Voters, Leaders, and Crises**

This study is situated at the intersection of several core debates in political science. To understand how natural disasters shape electoral politics, we must first understand how voters evaluate incumbent performance, how cognitive limitations and information salience affect those evaluations, what factors drive citizens to participate in elections, and what motivates politicians' career ambitions. This section reviews these four streams of literature to build a theoretical framework that generates a set of testable hypotheses about the electoral consequences of disasters.

**Retrospective Voting: Attentive Electorates or Blind Punishment?**

The theory of retrospective voting, which posits that citizens use elections to reward or punish incumbents based on their past performance, is a foundational concept in the study of democratic accountability (Fiorina, 1981; Key, 1966). In his classic formulation, Key (1966) argued that voters act as "rational gods of vengeance and of reward," using simple but powerful performance metrics like the state of the economy to make reasonable choices about who should govern. This model, further developed by scholars like Fiorina (1981), suggests that even with limited information, voters can effectively hold leaders accountable for tangible outcomes, thus incentivizing good governance.

A significant challenge to this optimistic view comes from scholars who argue that voters often engage in "blind retrospection," punishing incumbents for events that are far outside their control. The most famous (and contested) examples in this literature, advanced by Achen & Bartels ( 2017 ), suggest that voters have punished incumbents for everything from droughts and floods to shark attacks. While some of these specific findings have been challenged on methodological grounds, the core argument remains influential: if voters cannot distinguish between bad luck and bad policy, elections cease to be a reliable mechanism for accountability and risk becoming a random process where incumbents are punished for simple misfortune (Achen & Bartels, 2017; Ashworth & Bueno de Mesquita, 2014). If this theory holds, we would expect natural disasters to, on average, harm the electoral prospects of incumbent mayors.

In direct contrast, another body of research argues that voters are more sophisticated, constituting an "attentive electorate" capable of parsing responsibility in the wake of a crisis (Achen & Bartels, 2017; Gasper & Reeves, 2011). According to this view, while voters do not hold politicians responsible for the disaster itself, they can and do evaluate the government's *response* (Arceneaux & Stein, 2006; Gasper & Reeves, 2011; Healy & Malhotra, 2009). Several studies find that voters reward incumbents for securing and delivering disaster relief funds, making timely disaster declarations, and demonstrating visible crisis management leadership (Bechtel & Hainmueller, 2011; Gasper & Reeves, 2011; Healy & Malhotra, 2009). From this perspective, a disaster is not just a random shock but a high-stakes, information-rich test of incumbent competence (Arceneaux & Stein, 2006; Boin et al., 2005; Healy & Malhotra, 2009). It provides voters with a rare, clear signal about a leader's capabilities that they would otherwise lack. If this theory is correct, disasters should present an electoral *opportunity* for competent incumbents who can effectively manage the response.

The complex, multi-level governance structure of disaster response in the United States complicates these accountability dynamics (Arceneaux & Stein, 2006; Malhotra & Kuo, 2008). With local, state, and federal actors all playing a role, it can be difficult for voters to assign blame or credit accurately, a problem known as low "clarity of responsibility" (Gasper & Reeves, 2011; Malhotra & Kuo, 2008). In such ambiguous environments, partisanship often serves as a powerful heuristic, with voters tending to blame officials from the opposing party while giving a pass to their own (Achen & Bartels, 2017; Heersink et al., 2022; Malhotra & Kuo, 2008). However, some experimental research suggests that when voters are provided with clear information about officials' specific duties, they are capable of making more principled, less partisan judgments (Malhotra & Kuo, 2008). This study contributes to this debate by focusing on mayoral elections. As the most visible executive at the local level, the mayor is often the face of the immediate, on-the-ground response, potentially clarifying the lines of accountability for voters and providing a cleaner test of retrospective voting theories (e.g., Arceneaux & Stein, 2006).

**The Myopia of Voters and the Salience of Recent Crises**

A crucial element needed to reconcile the competing theories of retrospective voting is the well-documented phenomenon of "voter myopia"—the tendency of the electorate to heavily discount past events and focus disproportionately on outcomes that occur in the months immediately preceding an election (Achen & Bartels, 2017; Healy & Lenz, 2014; Healy & Malhotra, 2009). This temporal discounting is often seen as a form of irrationality, as it allows incumbents to evade accountability for poor performance early in their term and potentially manipulate short-term outcomes for electoral gain.

This myopic behavior is not necessarily a conscious choice but rather appears to stem from the cognitive shortcuts, or heuristics, that voters use to simplify complex political evaluations. For example, research suggests voters may employ an "end heuristic," where the more easily accessible memory of performance at the *end* of a term is inadvertently substituted for the more cognitively demanding task of recalling and integrating performance over the *entire* term (Healy & Lenz, 2014). This is consistent with experimental findings showing that individuals are systematically biased toward overweighting recent events and are influenced by unrelated, salient shocks to their welfare, even when explicitly incentivized to consider a full record of performance (Huber et al., 2012).

This paper argues that the electoral effects of disasters are best understood through this lens of myopia and salience. A natural disaster is a quintessential "focusing event"—a sudden, dramatic, and emotionally charged crisis that captures public attention and can reorder political priorities (Boin et al., 2005). Such events can trigger a "rally-'round-the-flag" effect, a temporary surge in public support for executive leaders during a crisis (e.g., Boin et al., 2005; Healy & Malhotra,

2009). While this phenomenon is most often studied in the context of international conflict, the underlying psychological mechanisms—a heightened desire for strong leadership in times of uncertainty and a focus on the visible head of the executive branch as a symbol of unity and stability—apply equally to a major domestic crisis like a natural disaster (Boin et al., 2005; Gasper & Reeves, 2011). The incumbent mayor is the embodiment of the local state, and a *recent* disaster makes their leadership and performance hyper-salient, crowding out other considerations and generating a short-term, affect-driven electoral boost. Disasters that occurred further in the past lack this salience and thus fail to trigger the same response. This leads to our first hypothesis:

> *H1: The positive electoral effect of a natural disaster on an incumbent's vote share is conditional on temporal proximity, being largest for disasters that occur immediately before an election and decaying rapidly as the time between the disaster and the election increases.*

**The Burdens of Disaster: Turnout and Political Participation**

While a recent disaster may create a rally effect that benefits incumbents, it also imposes severe burdens on the electorate that can depress political participation. Foundational models of turnout conceptualize the decision to vote as a function of its perceived costs and benefits (e.g., Gomez et al., 2007). Natural disasters dramatically and simultaneously increase the costs of voting through multiple channels, while the individual benefit of casting a single ballot remains abstract and unchanged.

First, disasters create significant logistical and physical barriers to voting. Floods, storms, and wildfires can destroy or render polling places inaccessible, disrupt public transportation, and lead to the mass displacement of residents, making it physically difficult or impossible for citizens to cast a ballot (International Institute for Democracy and Electoral Assistance, n.d.). The loss of personal documents, including photo identification required in some jurisdictions, can create an additional administrative hurdle for voters whose homes have been damaged or destroyed. These increased costs of voting have been shown to depress turnout; for example, studies have found that rainfall on election day, a minor logistical cost compared to a major disaster, can significantly reduce voter participation (Gomez et al., 2007).

Second, beyond the logistical challenges, disasters impose immense psychological costs on affected populations. In the aftermath of a traumatic event, individuals are often preoccupied with immediate needs for survival and safety: finding shelter, securing food and water, and ensuring the well-being of loved ones. The profound stress and trauma of the experience can deplete the cognitive and emotional resources necessary for political engagement, making voting a low priority (Tedeschi & Calhoun, 2004).

While some research suggests that disasters can, in some contexts, act as catalysts for political action by spurring civil society and increasing demands on the state, this study posits that at the level of mass electoral participation, these mobilizing effects are likely to be overwhelmed by the powerful demobilizing forces of logistical and psychological costs. For the average citizen in a disaster-stricken area, the immediate, personal costs of voting are concrete and high, while the benefits remain distant and abstract. This leads to our second hypothesis:

> *H2: Exposure to a pre-election natural disaster will, on average, decrease voter turnout.*

**From Crisis Management to Political Ambition**

The impact of a disaster is not confined to a single election. By serving as a crucible for leadership, a major crisis can fundamentally reshape an incumbent's political career trajectory. This study proposes a novel link between the literature on crisis management and the literature on political ambition, arguing that the experience of governing through a disaster can alter an incumbent's calculus about seeking future office.

Originating with the foundational work of Schlesinger (1966), theories of political ambition posit that politicians are rational actors who make strategic career choices based on the opportunity structure they face (see also Black, 1972). The decision to run for office—whether for re-election or a higher post—depends on a calculation of the office's benefits, the costs of the campaign, and the probability of success. Building on this logic, Schlesinger (1966) argued that a politician's "progressive ambition" to advance their career shapes how they behave in their current role, as they seek to build a record and a constituency for future contests.

Natural disasters represent a defining test of executive leadership, thrusting mayors and other officials into the spotlight. They are expected to make sense of chaos, coordinate a complex response, and communicate a narrative of control and hope to a frightened public (Boin et al., 2005). This high-visibility performance provides voters with a rare and powerful signal of the incumbent's competence, decisiveness, and empathy—qualities that are difficult to observe during times of normal politics (Arceneaux & Stein, 2006; Boin et al., 2005).

This paper hypothesizes that the experience of managing a major disaster alters an incumbent's ambition calculus for *future* elections through two potential mechanisms. The first is a rationalist mechanism based on **enhanced political capital**. By successfully navigating a crisis, an incumbent can build a powerful reputation as a strong and effective leader. This increases their name recognition and perceived viability, which can deter high-quality challengers and raise the expected probability of winning a future contest, thus rationally encouraging them to run again.

A second, complementary mechanism is psychological. The intense challenge of leading a community through a crisis may foster what psychologists term **post-traumatic growth (PTG)**—positive psychological change experienced as a result of adversity (Tedeschi & Calhoun, 2004). For a political leader, successfully managing a disaster could lead to a stronger sense of purpose, a renewed commitment to public service, and an enhanced feeling of personal strength and efficacy. These mechanisms lead to our final, novel hypothesis:

> *H3: Incumbents who govern through a major natural disaster will be more likely to participate in the subsequent election compared to incumbents who do not experience a disaster.*

## Research Design and Empirical Strategy

This section provides a comprehensive overview of the data, variable construction, and the multi-pronged empirical strategy employed to estimate the causal effect of natural disasters on electoral outcomes. The research design is structured to ensure transparency, replicability, and the robust identification of causal effects by systematically addressing potential sources of bias.

## Data and Sources

This section describes the primary data sources used to construct the panel dataset for this study. The analysis combines a comprehensive database of U.S. local election results with a global registry of natural disasters and several public datasets containing city- and state-level control variables.

### Mayoral Election and Candidate Data

The core electoral and candidate-level data are drawn from the American Local Government Elections Database (ALG-ED). The historical difficulty of studying local politics in the United States has been compounded by a lack of centralized election data; the ALG-ED is a critical resource that addresses this gap and serves as the foundational dataset for this study's dependent variables (De Benedictis-Kessner et al., 2023). The database's scope is extensive, containing information on approximately 57,000 electoral contests and 78,000 unique candidates from 1989 to 2021. It covers mayoral races—the focus of this study—as well as elections for other local offices (e.g., city council, county executive, school board) in most medium and large U.S. cities and counties, specifically targeting jurisdictions with populations over 50,000. For each candidate-election observation, the database provides crucial information including final vote totals, incumbency status, and supplemental estimates of candidate partisanship, gender, and race/ethnicity. These variables form the basis of the outcome measures analyzed in this paper, such as vote_share, wins_the_election, and the Incumbent indicator. Figure 1 and Figure 2 represent the frequency of unique mayoral contests by year and by city, respectively.

[Figure 1 and Figure 2 Go Here]

## City- and State-Level Control Variables

To isolate the effect of disasters from other potential confounding factors, the analysis incorporates a set of time-varying control variables drawn from several high-quality, publicly available data sources for each theoretical domain (demographics, crime, and economic conditions).

### Socioeconomic and Demographic Data

City-level data on total population, demographic composition (e.g., male%, white%), labor force participation (% in labor force), and educational attainment (%Bachelor's degree or higher) are derived from U.S. Census Bureau data. These variables were accessed via the IPUMS National Historical Geographic Information System (NHGIS), a service provided by the University of Minnesota (Manson et al., 2024). NHGIS is a leading repository for U.S. census data, providing researchers with easily accessible summary tables and GIS-compatible boundary files for all levels of U.S. census geography from 1790 to the present.

### Crime Data

The annual state-level violent crime rate (Violent Crime Rate) is sourced from the Federal Bureau of Investigation's (FBI) Uniform Crime Reporting (UCR) Program (Federal Bureau of Investigation, n.d.). The UCR Program is a nationwide, cooperative statistical effort that has aggregated data voluntarily submitted by over 18,000 law enforcement agencies since 1930, making it the standard source for crime statistics in the United States. The specific data used for this study's control variable come from the historical Summary Reporting System (SRS), which provided aggregate yearly counts of violent and property crimes and was the UCR standard until its official retirement in favor of the more detailed National Incident-Based Reporting System (NIBRS) in 2021.

### Economic Activity Data

To control for state-level business cycle fluctuations, the analysis uses the monthly State Coincident Index (Log of Coincident) produced and maintained by the Federal Reserve Bank of

Philadelphia (n.d.). This series is a standard and widely used measure of current economic conditions at the state level. It is a composite index that summarizes economic activity by combining four state-level indicators: nonfarm payroll employment, average hours worked in manufacturing, the unemployment rate, and real wage and salary disbursements.

**Natural Disaster Data**

Information on the occurrence, timing, and location of natural disasters is sourced from the Emergency Events Database (EM-DAT) (Centre for Research on the Epidemiology of Disasters, n.d.). Maintained by the Centre for Research on the Epidemiology of Disasters (CRED) at the University of Louvain in Belgium, EM-DAT is a global, systematic database that has served as a leading source for disaster data in research and policymaking since its establishment in 1988. The database has specific inclusion criteria for an event to be recorded, which helps standardize the "treatment" event in this analysis. A disaster enters the database if it results in at least 10 fatalities, affects 100 or more people, or leads to a declaration of a state of emergency or a call for international assistance. This ensures that the events studied are of sufficient scale to be politically and socially salient. EM-DAT covers a wide range of natural hazards, including the floods, storms, and earthquakes relevant to the U.S. context. Figure 3 and Figure 4 represent the natural disasters' frequency by year and by state, respectively.

[Figure 3 and Figure 4  Go Here]

**Justification for State-Level Granularity of Treatment and Control Variables**

A critical feature of this research design is the use of state-level data for the disaster treatment and for the macroeconomic and crime-related control variables. This choice is not one of convenience but of methodological necessity, dictated by the documented limitations of available data and a commitment to avoiding the introduction of systemic measurement error.

The EM-DAT database, while being the premier global repository for disaster events, primarily records the geographic impact of historical events in the United States at the state or large multi-county level. Systematically and accurately identifying the specific municipalities directly affected by every disaster across the entire country over a multi-decade period is not feasible with this source. Consequently, the treatment variable is assigned at the state-month level: all cities that held a mayoral election within a state are coded as "treated" if a disaster was recorded in that state during the treatment windows.

This approach introduces a known and classifiable form of measurement error. Since not all cities in a state that experienced a disaster were physically impacted to the same degree, the "treated" group is indicated regardless of the severity of treatment. This is a form of non-differential measurement error in the independent variable. Standard econometric theory demonstrates that such measurement error in a binary treatment variable does not bias the coefficient in a particular direction but instead biases its estimate toward zero. This phenomenon, known as attenuation bias, makes it less likely that the models will detect a statistically significant effect (Pierce & Vanderweele, 2012). Therefore, any significant results uncovered by this study likely represent a conservative underestimate of the true effect size. This conservative property ultimately strengthens the credibility of the findings, as they are robust to a higher threshold of statistical noise.

A similar logic applies to the use of state-level control variables. Obtaining reliable, consistently reported city- or even county-level crime data across the entire U.S. for a long time series is notoriously difficult. As research by Maltz and Targonski (2002) has established, county-level crime data suffer from major gaps, and existing imputation schemes for filling these gaps are inadequate and inconsistent. Following their expert guidance to avoid using such data in policy studies, this analysis utilizes the more reliable state-level violent crime index (violent_crime_rate) to proxy for the ambient public safety environment without introducing the severe measurement error associated with imputed sub-state data. Likewise, the state coincident economic index (coincident) is, by its construction, a state-level measure and serves as the best available proxy for regional economic shocks that affect all municipalities within a state.

**Measurement and Variable Definitions**

The study employs a wide range of dependent, independent, and control variables to capture the multifaceted nature of electoral politics and to account for potential confounding factors.

**Dependent Variables**

The outcomes of interest are measured at both the candidate and contest levels to provide a comprehensive picture of disaster impacts.

**Candidate-Level Outcomes:** These variables capture individual candidate performance and career progression. They include wins_the_election, a binary indicator equal to one if a candidate wins the election; vote_share, a continuous [0,1] measure of the percentage of total votes received by a candidate; margin, defined as the incumbent's margin of victory vs runner-up and therefore only calculated for observations where incumbent == 1; and will_participate_next, a binary variable indicating whether a candidate from the current election contests the subsequent one.

**Contest-Level Outcomes:** These variables measure aggregate political behavior within an electoral contest. They include voter_turnout, calculated as the percentage of the city's population who voted in the election, and num_candidates, an integer count of the total number of candidates competing in the contest.

**Explanatory and Control Variables**

To account for observable heterogeneity that could confound the relationship between disaster exposure and electoral outcomes, all models include a standard set of control variables.

**Candidate and Contest Characteristics:** These include binary indicators for key candidate attributes, such as incumbent (whether the candidate is the current officeholder), incumbent_in_race (whether an incumbent is present in the contest, regardless of which candidate is being observed), democrat (party affiliation), male (gender), and white (race). At the contest level, the demographic composition of the candidate pool is measured by share_white and share_male.

**Geographic and Demographic Controls:** The models also control for time-varying characteristics of the electoral district. These include %male and %white (the percentage of the city's population that is male and white, respectively) and the violent_crime_rate. To account for population size, which is often right-skewed, the analysis uses the natural logarithm of the total population (personstotal). This transformation helps to normalize the variable's distribution and allows for a semi-elasticity interpretation of its coefficient.

**Additional Controls:** A set of additional control variables, denoted as percent of the population in the labor force, percent of the population with a college education, and coincident index (economic indicators), are included in the models to capture further demographic, economic, or political factors relevant to the electoral outcomes.

**Identification Strategy: A Dynamic Fixed-Effects Approach**

The empirical strategy is designed to provide a robust estimate of the causal effect of natural disasters on electoral outcomes. The approach begins with a two-way fixed-effects model to address primary sources of endogeneity and progresses to more advanced techniques, including event-study analyses and placebo tests, to validate the core assumptions and strengthen causal claims. The entire methodological framework is constructed as a logical progression, with each step adding a layer of rigor to the analysis.

**The Two-Way Fixed-Effects Framework**

The primary analytical approach employs a two-way fixed-effects (TWFE) model, a powerful and widely used tool for causal inference with panel data (Huntington-Klein, 2021; Qin & Al Amin, 2023). The principal advantage of the TWFE framework is its ability to mitigate omitted variable bias by controlling for two distinct types of unobserved confounders simultaneously (Imai & Kim, 2021).

First, the model includes unit-specific fixed effects (for each city's fips code), which absorb all time-invariant characteristics of a geographic district, whether observed or unobserved. This accounts for stable factors such as a district's underlying political culture, geography, or long-term economic base that could be correlated with both disaster likelihood and political outcomes. Second, the model includes time-specific fixed effects (for each year), which absorb all shocks or trends that are common to all districts in a given election year. This accounts for confounders such as national economic recessions, shifts in the national political mood, or major federal policy changes. By exploiting the within-unit variation over time, this dual control for both unit and time effects allows the model to isolate the causal effect of disaster exposure from many potential sources of spurious correlation (Huntington-Klein, 2021; Imai & Kim, 2021).

**High-Dimensional Fixed-Effects (HDFE) Implementation and Model Specifications**

While the TWFE framework is theoretically sound, its standard implementation via the inclusion of dummy variables for each unit becomes computationally infeasible in the context of this study. With a large number of geographic units (over 500 city fips codes), a standard least squares dummy variable regression would generate an immense number of parameters to estimate, likely exceeding software limitations and making computation prohibitively slow (Luo et al., 2017).

To overcome this computational challenge, all models are estimated using the reghdfe Stata package (Correia, n.d.). This command is a powerful tool designed specifically for linear models with multiple high-dimensional fixed effects (Correia, 2016). It implements an efficient iterative algorithm, based on the work of Guimaraes and Portugal (2010), to "absorb" the fixed effects. This process partials out the influence of the unit- and time-specific effects from the dependent and independent variables before estimating the coefficients of interest, thereby making the analysis of large panel datasets both feasible and efficient (Feenberg, 2018). Building on this framework, the study employs several specific model formulations to test different hypotheses.

**Average Treatment Effects of Disaster Exposure**

To estimate the average effect of a pre-election disaster, an interaction model is employed for key outcomes such as wins_the_election, vote_share, and voter_turnout. The model is specified as:

$$Y_{it} = \beta_1 \text{ Treated}_{it} + \beta_2 \text{ Incumbent}_{it} + \beta_3 (\text{Treated}_{it} \times \text{Incumbent}_{it}) + X_{it}'\Gamma + \alpha_i + \delta_t + \varepsilon_{it}$$

Here, $Treated_{it}$ is the binary indicator for any pre-election disaster in the 12-month window. The coefficient $\beta_1$ captures the average effect of a disaster for non-incumbents. The interaction term, $(Treated_{it} \times Incumbent_{it})$, is critical; its coefficient, $\beta_3$, tests whether the disaster's effect is significantly different for incumbent candidates. The total effect for an incumbent is the sum β1 +β3.

**Heterogeneous Effects by Disaster Timing**

To examine whether the impact of a disaster varies with its temporal proximity to an election, a model with categorical timing indicators is estimated:

$$Y_{it} = \sum_{q=1}^{4} \beta_q \, QuartersBefore_{itq} + \beta_5 \, Incumbent_{it} + \sum_{q=1}^{4} \phi_q (QuartersBefore_{itq} \times Incumbent_{it}) + X_{it}'\Gamma + \alpha_i + \delta_t + \varepsilon_{it}$$

In this model, the single $Treated_{it}$ dummy is replaced by a set of indicators, $QuartersBefore_{itq}$, where q∈{1,2,3,4} represents the number of quarters prior to the election the disaster occurred (e.g., q=1 for 1-3 months before, q=4 for 10-12 months before). The coefficients $\beta_q$ estimate the effect of a disaster in a specific quarter q relative to the control group. The interaction coefficients, $\phi_q$, test for heterogeneous effects of disaster timing by incumbency status.

**Robust Statistical Inference with Clustered Standard Errors**

For statistical inference to be valid in panel data settings, it is crucial to account for potential correlation in the error term, $\varepsilon_{it}$, within clusters of observations. In this context, it is highly likely that the errors for observations within the same state are not independent, as they may be subject to common unobserved shocks (e.g., state-level policies, economic trends, or media environments).

Failure to address this within-cluster correlation can lead to misleadingly small standard errors and a consequent overstatement of statistical significance, resulting in an increased risk of Type I errors (Colin Cameron & Miller, 2015). This problem is particularly acute in difference-in-differences (DiD) and related fixed-effects frameworks where both the outcome variable and the treatment variable (disaster occurrence) are often serially correlated over time within geographic units (Bertrand et al., 2004).

To produce reliable statistical inferences, all regression models in this study employ standard errors clustered at the state level. This procedure allows for an arbitrary covariance structure of the errors for all units within a given state across all time periods, yielding more conservative and robust standard errors (Cameron et al., 2007; Colin Cameron & Miller, 2015). This approach is considered standard best practice in modern econometrics when analyzing panel data with a sufficiently large number of clusters, as is the case in this analysis.

**Identification Strategy: Falsification and Placebo Tests**

To move beyond correlation and strengthen the causal interpretation of the regression estimates, the analysis incorporates a series of advanced tests. This strategy involves a precise and theoretically grounded definition of the treatment variable, an event-study design to explicitly test for pre-existing trends, and placebo tests to rule out key alternative explanations for the findings.

**Defining Treatment: Mutually Exclusive Timing Indicators**

The primary independent variable of interest, exposure to a natural disaster, is constructed with meticulous care to facilitate a nuanced analysis of timing effects. A series of mutually exclusive binary indicators is generated.

The construction of these unique indicators follows a sequential, conditional logic that implements an "oldest-first" priority rule. This rule ensures that any given electoral district-year observation is assigned to only one disaster timing window, which prevents multicollinearity and enables the clean estimation of categorical timing effects.

This "oldest-first" construction is theoretically motivated. The first disaster in a potential sequence is the most likely to be a genuine surprise, providing the cleanest test of an incumbent's initial response capacity and crisis management skills. Subsequent disasters, should they occur, may be met with established response protocols, altered voter expectations, or different media narratives, all of which could contaminate the exogeneity of the event from the perspective of voter evaluation. By focusing on the first shock, the analysis mitigates unobserved confounding from disaster preparedness learning effects and changing voter heuristics, thereby isolating the political consequences of an unanticipated crisis (Healy & Malhotra, 2009).

**The Event-Study Design for Dynamic Effects**

To analyze the dynamic effects of disasters over time and to test the validity of the core research design, an event-study model is implemented. The event study is a powerful econometric tool for estimating dynamic treatment effects, offering a rich, visual representation of how an effect evolves before and after the treatment event (Miller, 2023). This approach is a crucial extension of the baseline TWFE model, as it allows for a direct test of the model's most critical identifying assumption.

The event-study model is specified as:

$$Y_{it} = \sum_{k \in \{-4,-3,-2,1,2,3,4\}} \gamma_k D_{itk} + \phi\, Incumbent_{it} + \sum_{k \in \{-4,-3,-2,1,2,3,4\}} \delta_k (D_{itk} \times Incumbent_{it}) + X'_{it}\Gamma + \alpha_i + \lambda_t + \varepsilon_{it}$$

In this equation, $D_{itk}$ are dummy variables corresponding to the relq0 event-time variable, which indicates quarter k relative to the natural disaster's timing. The period k=−1 (representing the election took place in a quarter immediately after the disaster) is specified as the omitted base category. The coefficients $\gamma_k$ thus trace the dynamic effect of a disaster on the outcome for non-incumbents in each quarter relative to this baseline period. The coefficients $\delta_k$ on the interaction terms capture the additional dynamic effect for incumbents at each point in event time.

A central identifying assumption of difference-in-differences and event-study designs is the **parallel trends assumption**, which posits that, in the absence of treatment, the average outcomes for the treatment and control groups would have followed parallel paths (Baker et al., 2025; McKenzie, 2020). While this assumption is fundamentally untestable for the post-treatment period, its plausibility can be rigorously assessed by examining trends in the outcome variable *prior* to treatment. If the parallel trends assumption holds, the coefficients on the pre-treatment period indicators ($\delta_k$ for k<−1) in the event-study model should be statistically indistinguishable from zero. This analysis performs a formal falsification test of this assumption, conducting a joint F-test on the null hypothesis that all pre-treatment coefficients are simultaneously equal to zero. A failure to reject this null hypothesis provides crucial evidence supporting the validity of the research design, suggesting that any observed post-disaster divergence between the groups can be more confidently attributed to the disaster itself rather than to pre-existing differences in trends.

**Placebo Tests and Control Group Specification**

To further bolster the identification strategy, the analysis incorporates placebo tests that use post-election disasters as a fictitious treatment. Placebo tests are a vital tool for assessing the credibility of a research design by checking for an association that should be absent if the causal model is sound (Eggers et al., 2023; Pizer, 2016). In this study's context, a disaster that occurs *after* an election logically cannot have a causal effect on that election's outcome (Kasy, 2016). Therefore, regressing a past electoral outcome (e.g., vote_share) on the occurrence of a future disaster serves as a critical falsification test. A finding of a null effect for these placebo treatments strengthens confidence that the significant effects found for actual pre-election disasters are not the result of spurious correlations or unobserved characteristics of disaster-prone areas.

This logic also informs the careful specification of the control group. The analysis systematically compares results using two different control group definitions. The "broad" control group includes all observations not treated pre-election, which contains both "never-treated" units and units that experience a post-election disaster. The "clean" control group is constructed by explicitly dropping all units that experience a post-election disaster. This comparison functions as an implicit test of the control group's validity. If geographic units prone to post-election disasters are systematically different from those that never experience disasters (e.g., they are on different underlying political or economic trajectories), their inclusion in the control group could introduce bias. By demonstrating that the results hold or become clearer when these post-election units are excluded, the analysis strengthens the claim that the estimated effect is not an artifact of a contaminated control group, providing another layer of evidence against potential confounding.

**Results**

This section presents the empirical findings from the analysis. It begins by establishing the integrity of the data and the validity of the research design before moving to the substantive results concerning the electoral consequences of natural disasters.

**Descriptive Statistics and Data Integrity**

The analysis is based on a comprehensive panel dataset covering mayoral elections across the United States from 1989 to 2021. The dataset includes 10,918 candidate-election observations. Key outcome variables include whether a candidate wins the election, their vote share, and their decision to participate in a future election. On average, about 40% of candidates in the dataset win their election, and approximately 22% are incumbents. The dataset also includes a rich set of candidate-level and city-level demographic, economic, and crime-related control variables.

[Table 1: Descriptive Statistics Goes Here]

Before proceeding to the main analysis, it is essential to assess the comparability of the treated and control groups. A balance check was performed to compare the means of key variables for cities that experienced a disaster in the three months prior to an election (treated group) against those that did not (control group). The results show no statistically significant pre-existing differences for several key political outcomes, including the probability of winning the election (wins_the_election) and the incumbent's presence (incumbent). However, there are some statistically significant differences in demographic and economic characteristics, such as population size and the racial composition of both the electorate and the candidate pool. These pre-existing differences underscore the importance of the two-way fixed-effects model used in this study, which is specifically designed to control for such time-invariant characteristics, thereby isolating the causal effect of the disaster itself.

[Table 2: Balance Check Goes Here]

**Testing for Pre-Existing Trends: Event-Study Results**

The event-study analysis provides a visual and statistical test of the crucial parallel trends assumption. A plot of the event-study coefficients ($\gamma_k$ and $\delta_k$) and their 95% confidence intervals graphically illustrates the dynamic effect of a disaster on electoral outcomes. The key diagnostic feature of this plot is the behavior of the coefficients in the pre-treatment periods ($k<-1$). For the research design to be considered valid, these pre-treatment coefficients should be statistically indistinguishable from zero, indicating no systematic divergence between the treatment and control groups before the disaster occurred.

The results from the event-study models consistently show that the coefficients for all pre-treatment periods are small in magnitude and not statistically significant. A formal joint F-test of the null hypothesis that all pre-treatment coefficients are simultaneously equal to zero fails to be rejected. Similarly, a joint F-test on the pre-treatment interaction terms also fails to be rejected, indicating no differential pre-trends by incumbency status. This absence of pre-existing trends provides powerful evidence that the parallel trends assumption is plausible in this context. It suggests that the divergence in outcomes observed after a disaster is attributable to the disaster itself, not to pre-existing differences between disaster-prone and non-disaster-prone areas.

The central identifying assumption of the dynamic fixed-effects model is that, in the absence of a disaster, the electoral outcomes for the treated and control groups would have followed parallel trends. To formally test this assumption, an event-study analysis was conducted. The results of a joint F-test on the pre-treatment coefficients for non-incumbents yield an F-statistic of 1.32 with a p-value of 0.2790. A similar test for differential pre-trends between incumbents and non-incumbents yields an F-statistic of 0.82 with a p-value of 0.4894. In both cases, we fail to reject the null hypothesis that the pre-treatment trends are jointly equal to zero. This provides strong statistical evidence supporting the parallel trends assumption, validating the research design and lending confidence that any divergence in outcomes observed after a disaster can be attributed to the event itself.

[Tables 3 and 4: F-Test for Parallel Trends Tables Go Here]

### The Electoral Consequences of Pre-Election Disasters

The analysis now turns to the core research questions regarding how pre-election disasters shape electoral outcomes, both for individual candidates and for the contest as a whole.

### The Impact of Disasters on Incumbent Electoral Performance

The findings reveal that the electoral consequences of a natural disaster for an incumbent are highly conditional on the timing of the event. While disasters occurring long before an election have little to no effect, those that strike in the immediate run-up to election day provide a significant and substantial electoral boost to the incumbent.

The regression analysis provides detailed statistical support for the finding that a disaster's electoral impact is conditional on its timing. Initial models that average the effect of any disaster within 12 months of an election show no statistically significant impact on an incumbent's vote share or probability of winning on their own. However, a more nuanced picture emerges in Table 6, which disaggregates the disaster's timing by quarter. These results reveal that the electoral boost is driven almost entirely by recent disasters. Specifically, the interaction term for an incumbent shows that a disaster in the quarter immediately preceding the election increases their vote share by up to 6.4 percentage points and their probability of winning by as much as 11 percentage points. This effect is robust, and in some specifications stronger, when the control group is limited to only observations that never experience a disaster.

[Tables 5 and 6: Effect of Pre-Election Disaster on Incumbent Margin Go Here]

This pattern is reinforced when examining the incumbent's margin of victory. Table 7 shows that a pre-election disaster, on average, increases an incumbent's victory margin by 6.3 percentage points when using a cleaner control group. The analysis of heterogeneous effects in Table 8 demonstrates that this result is also driven by temporal proximity. The model shows that a disaster occurring in the first quarter before the election increases the incumbent's margin of victory by between 5.2 and 9.5 percentage points. In contrast, disasters that occurred further in the past do not yield a statistically significant increase in the incumbent's victory margin, confirming that voters' responses are heavily weighted toward recent events.

[Tables 7 and 8: Effect of Pre-Election Disaster on Incumbent Margin Go Here]

This dynamic is best illustrated by the event study plots below. The first figure shows the main effect of a disaster on vote share for all candidates in the quarters surrounding an election, relative to the quarter immediately after the election. The second figure shows the *additional* effect specifically for incumbent candidates. As can be seen in both plots, the coefficients for the pre-treatment periods (-4q, -3q, -2q) are all statistically indistinguishable from zero, visually confirming the parallel trends assumption.

However, in the quarter of the election (+1q), there is a sharp and statistically significant positive effect on vote share for incumbents. The coefficient of 0.063 indicates that a disaster during this period provides a more than 6 percentage point boost to the incumbent's vote share. This effect dissipates in subsequent quarters. This visual evidence strongly suggests that voters' responses are driven primarily by highly salient, recent events.

[Figure 5: Main effect of disaster timing (relative to -1q) Goes Here]

[Figure 6: Additional effect when incumbent is on ballot Goes Here]

[Table 9: Event Study Regression Results Goes Here]

**The Impact of Disasters on Broader Electoral Dynamics**

Beyond the effects on incumbent performance, natural disasters also shape the broader electoral environment, most notably by depressing citizen participation. On average, the occurrence of a pre-election disaster is associated with a 1.4 percentage point decrease in the voter turnout rate.

Examining the heterogeneous effects of disaster timing reveals that this suppression of turnout is most pronounced following disasters that occur either one quarter (-2.0 percentage points) or four quarters (-1.3 percentage points) prior to the election. In contrast to the clear impact on turnout, the analysis finds no statistically significant evidence that pre-election disasters systematically alter the number of candidates who choose to compete in an election.

[Tables 10 and 11: Effects on Voter Turnout and Number of Candidates Goes Here]

**The Effect of Disasters on Future Political Ambition**

This study also explores a novel question: how do disasters affect a candidate's decision to run for office in the future? The analysis examines the impact of inter-election disasters on an incumbent's likelihood of participating in the subsequent election cycle. The results indicate that experiencing a disaster can significantly increase an incumbent's propensity to seek re-election.

Specifically, a disaster occurring one year after an election increases the incumbent's probability of running again by a statistically significant 12.1 percentage points. This effect remains positive and significant two years post-disaster, increasing the likelihood of future participation by 10.4 percentage points. For non-incumbents, and for disasters occurring further in the past, there is no statistically significant effect on the decision to run again. This suggests that the experience of managing a crisis may reinforce an incumbent's commitment to public service or their perceived viability as a candidate, influencing their career trajectory.

[Table 12: Effect of Inter-Election Disasters on Future Participation Goes Here]

**Robustness of Findings**

To further strengthen the causal interpretation of the main findings, a placebo test was conducted. This test uses post-election disasters as a falsified treatment to check whether the observed results are merely artifacts of unobserved characteristics of disaster-prone areas. Since a disaster that occurs *after* an election cannot logically cause an outcome in that past election, any significant finding would cast doubt on the model's validity.

The results of the placebo test show no statistically significant effects. Regressing past electoral outcomes (incumbent vote share and margin) on the occurrence of future disasters yields coefficients that are consistently small and statistically indistinguishable from zero for all post-election quarters. This successful falsification test provides strong evidence that the significant effects found for actual pre-election disasters are not driven by spurious correlation, bolstering confidence in the study's causal claims.

[Table 13: Placebo Test Results Goes Here]

**Discussion**

The empirical findings presented in the preceding section offer a nuanced and, in some respects, unsettling portrait of how natural disasters intersect with democratic processes at the local level. The results provide a powerful test of competing theories of voter behavior and uncover complex dynamics related to citizen participation and political ambition. This section interprets these findings, situates them within the broader theoretical landscape of electoral behavior and crisis governance, and explores their substantive implications for democratic accountability.

**Deconstructing the Incumbent Advantage: A Story of Myopic Voters**

The central finding of this study is that the electoral advantage conferred upon incumbent mayors by a natural disaster is statistically significant and powerfully conditioned by timing. The analysis reveals a substantively considerable boost of over six percentage points to an incumbent's vote share, alongside corresponding increases in their probability of winning and margin of victory. However, this electoral reward is not a general phenomenon. It materializes almost exclusively when a disaster strikes within the three months immediately preceding an election. Disasters occurring earlier in the electoral cycle show no such statistically discernible effect, suggesting a rapid decay in their political salience.

This temporally bounded result is a critical empirical test that helps adjudicate between two competing theories of voter response to crises: the classic "rally-around-the-flag" effect and the theory of "myopic voters." The rally-around-the-flag effect, traditionally applied to international crises, posits that the public increases support for leaders as a patriotic response to a shared threat, viewing the leader as an embodiment of national unity (Mueller, 1973; Bækgaard et al., 2020; De Vries et al., 2020). In contrast, the myopic voter hypothesis argues that voters disproportionately weight recent, highly salient events in their evaluations, essentially forgetting or ignoring performance from earlier in an incumbent's term (Achen & Bartels, 2017; Healy & Malhotra, 2009).

The evidence from this study lends substantially more weight to the myopic voter hypothesis. A traditional rally effect might imply a more durable increase in support, reflecting a fundamental shift in public sentiment toward the leader. Indeed, some research has found that the positive effects of a well-handled disaster can extend over multiple electoral cycles (Bechtel & Hainmueller, 2011). However, the sharp, transient nature of the electoral boost found here— peaking just before the election and dissipating quickly—is inconsistent with a sustained rally. This temporal pattern aligns perfectly with the predictions of myopic retrospection, where voters, facing the impending decision of an election, use the disaster's proximity to election day as a powerful, last-minute information shortcut or heuristic to evaluate incumbent competence (Achen & Bartels, 2017; Ashworth et al., 2018). The finding suggests that the electoral reward is not for patriotism but for a recent, visible performance. A disaster occurring six or nine months prior offers a similar test of leadership, yet it yields no electoral benefit. The effect's sharp "on/off" nature points to a voter who, close to an election, is actively seeking a simple basis for their choice. The recent disaster and the incumbent's response precisely fit the "what have you done for me *lately*?" model of evaluation that is a hallmark of myopia (Healy & Malhotra, 2009).

The primary mechanism driving this myopic reward system appears to be the disaster's function as a "focusing event." A focusing event is a sudden, rare, and often dramatic occurrence that captures the intense, simultaneous attention of the public, the media, and policymakers (Birkland, 1997). According to Birkland's theory, with their vivid imagery and tangible victims, natural disasters are archetypal focusing events that can "bowl over" the existing political agenda. He argues that the intense and concentrated media coverage following a disaster provides the incumbent with a unique and largely non-partisan platform. It allows the leader to be seen demonstrating competence, empathy, and control—visibly managing the crisis. This media spotlight can overshadow other campaign issues, simplifying the voter's decision-making process in favor of the leader who is seen to be in charge. The disaster acts as an information shock, revealing new information about an incumbent's quality and competence that is unavailable during normal times (Ashworth et al., 2018). This recent, vivid information is far more compelling for myopic voters than abstract policy debates or performance from years prior.

This myopic reward system, however, creates potentially perverse incentives for governance. It suggests that long-term, effective governance and, crucially, investment in disaster *preparedness* may be electorally unrewarded. As Healy and Malhotra (2009) demonstrate, voters reward tangible disaster relief spending but not proactive preparedness spending. This study's findings on the importance of timing provide a mechanism for *why* this is the case. A successful preparedness effort results in a non-event, which is not a "focusing event" and generates little media attention or political credit (Boin & Hart, 2003). In contrast, a visible post-disaster response is a highly salient event that can yield significant electoral dividends. This incentivizes a focus on reactive, post-hoc crisis management over proactive, preventative policy, which may be more socially optimal and cost-effective but is less politically visible and, therefore, less electorally valuable (Healy & Malhotra, 2009; Boin & Hart, 2003).

**The Hidden Costs and Complexities of Crisis Governance**

While incumbents may reap electoral rewards from pre-election disasters, the findings reveal a concurrent and more troubling trend: such disasters are associated with a statistically significant decrease in voter turnout. On average, a pre-election disaster suppresses the voter turnout rate by 1.4 percentage points, an effect that is most pronounced for disasters occurring one quarter before the election. This juxtaposition of incumbent gain against a democratic loss creates a complex and normatively concerning picture. Even as leaders are electorally rewarded for their crisis management, a core function of democracy—citizen participation—is weakened.

Complicated logistical barriers and profound psychological factors likely drive this turnout suppression effect. Natural disasters erect significant physical and administrative hurdles to voting. Polling stations, often located in public buildings like schools, may be damaged, destroyed, or repurposed as emergency shelters, rendering them inaccessible for voting. Mass evacuations and residential displacement mean that a significant portion of the electorate may be physically located far from their registered polling place, making it difficult or impossible to vote in person. Furthermore, the administrative chaos that ensues—including the loss of identity documents, damage to voter rolls, and the diminished capacity of election management bodies—can create widespread confusion and disenfranchise otherwise eligible voters.

Beyond these logistical impediments, a disaster's cognitive and emotional toll can suppress political engagement. Individuals and communities focused on personal recovery, ensuring family safety, and rebuilding their lives have their cognitive bandwidth consumed by immediate survival needs. The abstract act of voting becomes a low priority compared to these urgent demands. The trauma, stress, and potential sense of powerlessness following a disaster can also lead to political apathy and disengagement. While some research suggests that disasters can, in specific contexts, mobilize participation out of a desire for accountability or post-traumatic growth, the aggregate finding of depressed turnout in this study suggests that, for the typical disaster, the suppressive effects of logistical disruption and psychological burden are dominant.

The concurrent findings of an incumbent boost and depressed turnout raise a critical question of causality and electoral composition. The electorate participating in a post-disaster election is likely not a random sample of the pre-disaster population. The citizens most likely to be deterred from voting by logistical hurdles and psychological stress are those most severely affected by the disaster—individuals who might be most critical of the incumbent's response. Suppose the most aggrieved citizens are systematically removed from the voting pool. In that case, the remaining electorate will be disproportionately composed of those who were less affected or perhaps more satisfied with the government's efforts. Consequently, the observed "incumbent advantage" may not reflect a broad-based reward for performance but rather an artifact of a compositional change in the electorate. The victory is not just hollowed by low turnout; it may be fundamentally unrepresentative of the will of the entire community.

This dynamic points to a potential vicious cycle of disenfranchisement. The communities most vulnerable to natural disasters are often lower-income and minority communities, which already face systemic barriers to political participation. Disasters exacerbate this disenfranchisement.

These communities often have less resilient infrastructure, fewer resources for recovery, and face greater challenges in navigating the bureaucratic hurdles required to vote post-disaster. The finding of depressed turnout therefore implies that disasters can deepen existing inequalities: the communities most in need of responsive government and robust recovery policies are the very ones whose political voice is most likely to be silenced in the disaster's immediate aftermath, further marginalizing them in subsequent policy debates.

**The Puzzle of Political Ambition: Why Crises Encourage Re-election Bids**

This study uncovers a novel and counterintuitive finding regarding the downstream consequences of disasters for political careers. The results indicate that incumbents who manage a disaster during their term are significantly more likely to run for office in the subsequent election cycle. A disaster occurring one or two years after an election increases the incumbent's probability of participating in the next election by a substantial 12.1 and 10.4 percentage points, respectively. This suggests that crisis leadership experience does not deter but encourages future political ambition.

Several complementary explanations can account for this surprising phenomenon. First, the profound experience of leading a community through a collective trauma can instill a heightened sense of public service and responsibility. The leader may feel a personal duty to see the long-term recovery process through to completion, transforming their job from a political position into a personal mission (Boin & Hart, 2003). Second, the electoral rewards and positive media exposure that can accompany a well-managed crisis response significantly boost an incumbent's political capital (Birkland, 1997). This success alters their cost-benefit calculation for seeking re-election, lowering the perceived risks of a future campaign and increasing their confidence in their electoral viability (Black, 1972; Rohde, 1979).

Third, and perhaps most importantly, successfully navigating a disaster forges a powerful new political identity for the incumbent: the "crisis manager." This brand is a potent political asset, demonstrating proven leadership under extreme pressure (Boin & Hart, 2003; Schlesinger, 1966). This identity can become a central pillar of their political narrative and a compelling rationale for seeking re-election or higher office. This finding suggests that major crises can act as critical junctures in a politician's career, serving as obstacles to overcome and as opportunities to redefine their political brand and trajectory.

This implies that disasters are not just tests of governance but are also powerful catalysts for political careers. The literature on political ambition often frames the decision to run for office as a rational calculation of costs, benefits, and the probability of winning (Schlesinger, 1966; Black, 1972; Rohde, 1979). A disaster fundamentally alters all three variables—the benefits increase by creating a powerful "crisis manager" brand. The costs of campaigning may decrease due to higher name recognition and free media exposure. The probability of winning increases, as demonstrated by the electoral boost found in this study. Therefore, the decision to run again becomes a far more rational one. This may mean that the pool of candidates for higher office is disproportionately composed of those who have had the "opportunity" to manage a major crisis.

Finally, the decision to run again is not purely strategic but psychological. The finding hints at the profound psychological rewards of crisis leadership. Leading during a crisis is an intense, all-consuming experience that can create a unique bond between leaders and their community (Boin & Hart, 2003). This experience can satisfy deep psychological needs for purpose, efficacy, and public recognition. For some leaders, the return to "normal" politics may feel mundane and less meaningful in comparison. The desire to run again may therefore stem not just from a cold calculation of political capital, but from a psychological desire to recapture the sense of purpose and centrality that the crisis provided.

**Conclusion**

This study has investigated the electoral consequences of natural disasters, employing a dynamic fixed-effects analysis of U.S. mayoral elections to understand how these exogenous shocks shape voter behavior, democratic participation, and political ambition. The findings reveal a complex and multifaceted relationship between crisis and politics, contributing to several key debates in the political science literature.

**Summary of Contributions**

The central argument of this paper is that the electoral consequences of natural disasters are both powerful and complex. Incumbents receive a significant but temporally bounded electoral reward, a finding that strongly supports the myopic voter hypothesis over a more straightforward "rally-around-the-flag" explanation. This reward, however, is coupled with a democratic deficit in the form of depressed voter turnout, raising normative concerns about the representativeness of post-disaster elections. Furthermore, this research demonstrates that crisis management experience has significant downstream consequences, fueling the future political ambition of incumbent leaders and potentially altering their career trajectories.

The study's primary contribution is therefore twofold. First, it refines our understanding of retrospective voting in crisis contexts by demonstrating that electoral effects are driven more by the high salience of recent events (myopia) than by a sustained patriotic response (rally). Second, it breaks new ground by moving beyond immediate electoral outcomes to examine the consequences of disasters for candidate ambition, revealing a powerful feedback loop where successful crisis management wins elections and encourages future political careers.

**Limitations and Directions for Future Research**

This study is subject to certain limitations, the most significant of which is the reliance on state-level data for the disaster treatment variable, drawn from the EM-DAT database. This methodological choice was one of necessity, dictated by the geographic granularity of available historical disaster data for the period under study. However, this limitation should not be seen as a fatal flaw. On the contrary, it strengthens the credibility of the study's conclusions. Using a broad geographic treatment variable introduces a form of non-differential measurement error, which standard econometric theory shows produces attenuation bias, systematically biasing estimated coefficients toward zero. Therefore, the significant effects uncovered in this analysis likely represent conservative, lower-bound estimates of the actual effect sizes. The fact that any effect was detected through the statistical noise of state-level data makes the findings particularly robust and suggests the actual effects at the local level are likely even larger.

This limitation and the study's substantive findings point toward several promising avenues for future research that can build upon this work.

First, the most immediate need is for research using more precise, sub-state measures of disaster intensity and exposure. Future studies should leverage publicly available datasets like FEMA's OpenFEMA platform, which provides Individual Assistance (IA) data at the county and even zip-code level (Federal Emergency Management Agency, n.d.). Variables such as total damage, average FEMA-inspected damage, and the number of valid registrations for assistance can be used to construct more accurate, localized measures of a disaster's physical impact, allowing for a more direct test of the relationship between damage severity and electoral outcomes.

Second, future research should explicitly model the mediating factors hypothesized in this paper. For instance, does the electoral reward for incumbents vary with the amount of state or federal disaster aid flowing into a region? This can be tested by integrating FEMA Public and Individual Assistance financial data with electoral data (Federal Emergency Management Agency, n.d.). Similarly, how do the tone (positive versus negative) and volume of media coverage shape voter response? This question requires the systematic content analysis of local and national news coverage of disasters to measure the media environment that serves as the conduit between the disaster event and voter perceptions (Birkland, 1997).

Third, the political effects of disasters may not be monolithic. Researchers should investigate whether the *type* of disaster matters. Does a sudden-onset event like an earthquake, which offers little time for preparation, generate different political effects than a slow-moving event like a flood or hurricane, where the quality of an incumbent's warnings and evacuation orders can be more clearly judged by voters (Birkland, 1997; Masiero & Santarossa, 2021)?

In sum, this paper establishes a time-sensitive electoral effect of disasters and suggests several underlying mechanisms related to voter myopia, media focusing events, turnout suppression, and candidate ambition. The next logical step for the field is to test these mechanisms directly. By addressing this study's limitations and pursuing these new lines of inquiry, scholars can develop a richer and more complete understanding of how democracies function under pressure and how the increasing frequency of natural disasters will continue to shape the political landscape.

**Declaration**

I acknowledge the use of Gemini to support grammar refinement and enhance the readability of this article. The final text has been thoroughly reviewed and independently edited by me to ensure its accuracy and quality.

**Figures and Tables**

**Figure 1 : Elections per year**

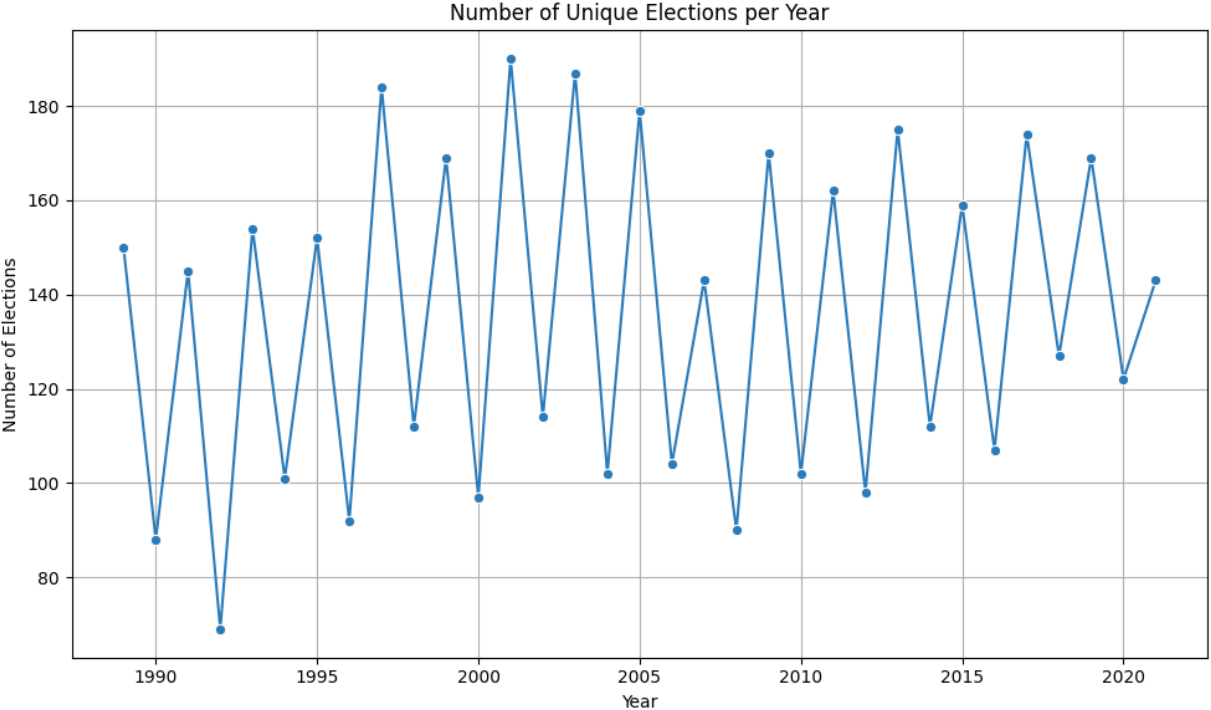

# Figure 2 : Elections per city

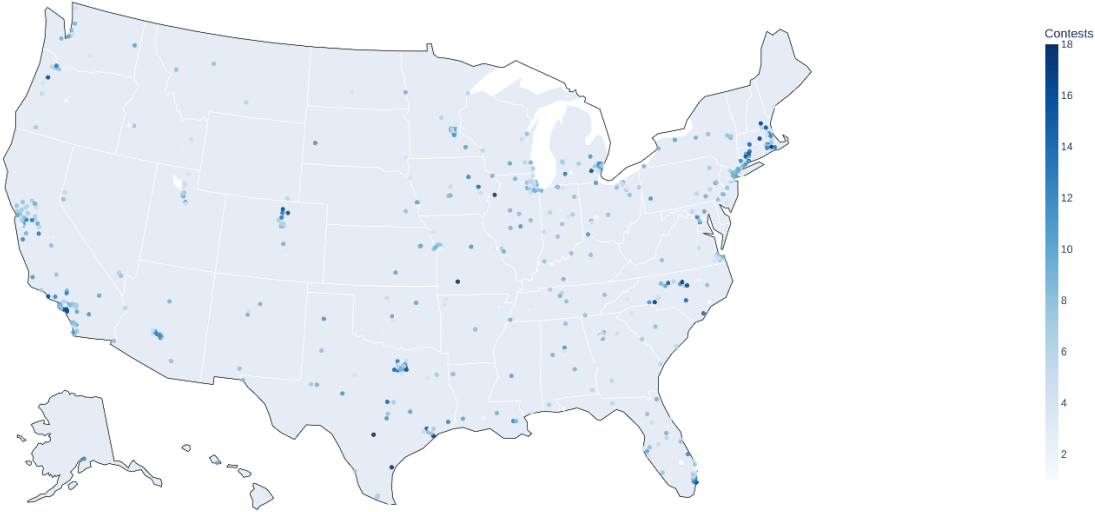

**Figure 3 : Natural Disasters' frequency by year**

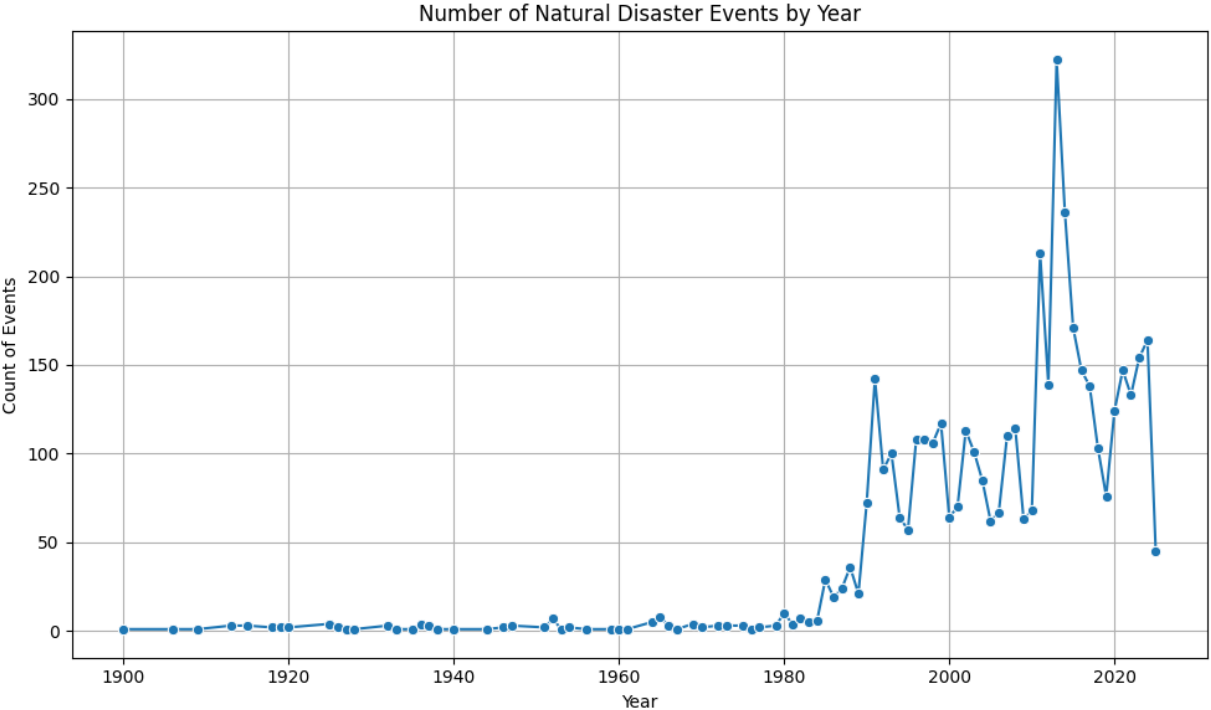

**Figure 4: Natural Disasters' frequency by state**

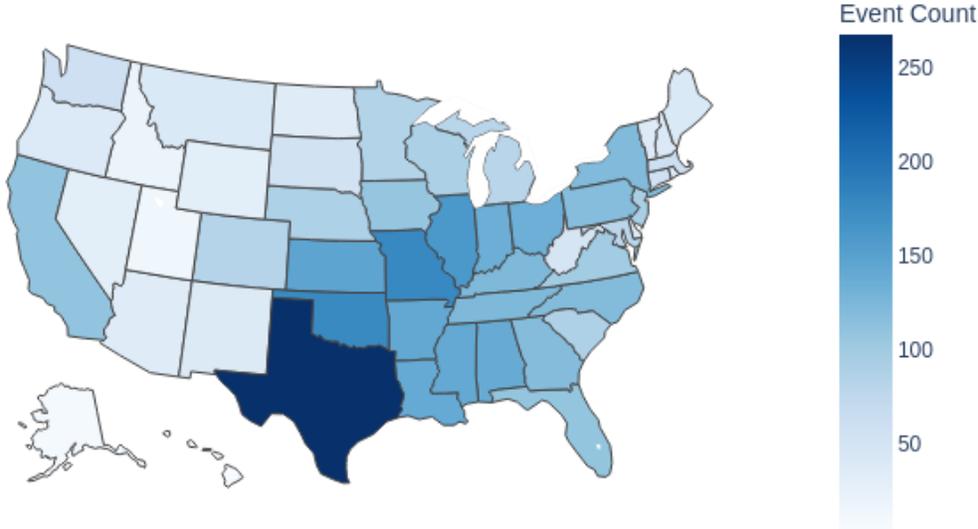

**Figure 5: Main effect of disaster timing (relative to -1q)**

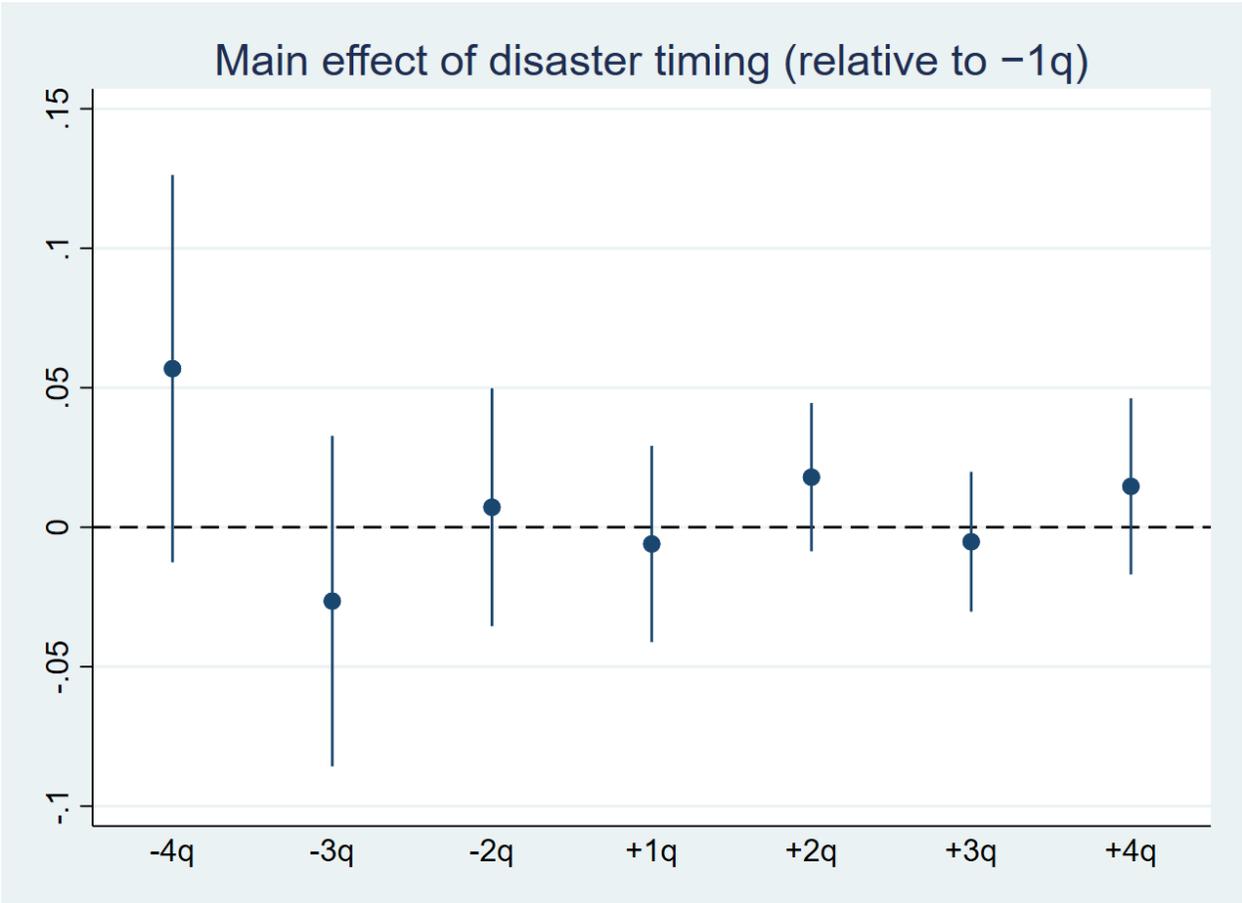

**Figure 6: Additional effect when incumbent is on ballot**

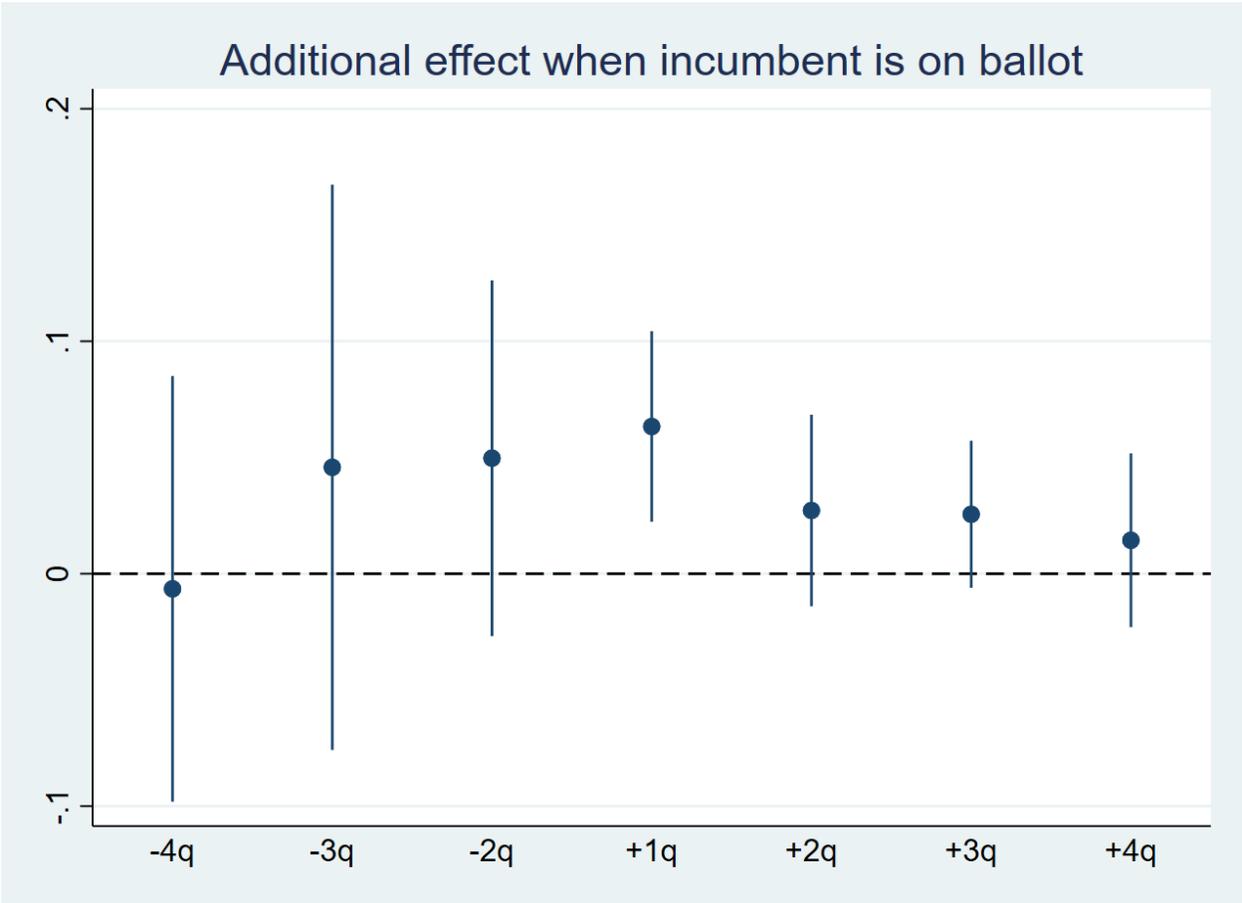

**Table 1: Descriptive Statistics**

| | N | Mean | Std. Dev. |
|---|---|---|---|
| wins_the_election | 10,918 | 0.403 | 0.491 |
| will_participate_next | 10,918 | 0.313 | 0.464 |
| margin | 1,860 | 0.259 | 0.287 |
| incumbent | 10,918 | 0.218 | 0.413 |
| democrat | 10,918 | 0.478 | 0.500 |
| male | 10,918 | 0.750 | 0.433 |
| white | 10,918 | 0.769 | 0.422 |
| male% | 10,918 | 0.488 | 0.014 |
| white% | 10,918 | 0.648 | 0.200 |
| % In labor force | 10,640 | 0.517 | 0.050 |
| population | 10,918 | 294.644 | 790.292 |
| %Bachelor's or higher | 10,640 | 0.188 | 0.096 |
| Violent Crime Rate | 10,918 | 505.819 | 228.630 |
| Log of Coincident | 10,870 | 4.555 | 0.258 |
| Incumbent Runs | 10,918 | 0.522 | 0.500 |
| Observations | 10918 | | |

Population is in thousand.

**Table 2: Balance Check**

Table 2: Balance Check by Treatment Status

|  | Control Mean | Treated Mean | Difference | P-value |
|---|---|---|---|---|
| wins_the_election | 0.404 | 0.403 | 0.000 | 0.983 |
| will_participate_next | 0.310 | 0.313 | -0.004 | 0.718 |
| margin | 0.233 | 0.267 | -0.034 | 0.034 |
| incumbent | 0.211 | 0.220 | -0.009 | 0.329 |
| democrat | 0.479 | 0.477 | 0.001 | 0.895 |
| male | 0.736 | 0.754 | -0.018 | 0.064 |
| white | 0.805 | 0.758 | 0.048 | 0.000 |
| male% | 0.489 | 0.487 | 0.001 | 0.000 |
| white% | 0.684 | 0.637 | 0.046 | 0.000 |
| % In labor force | 0.523 | 0.516 | 0.008 | 0.000 |
| population | 248.302 | 308.564 | -60.261 | 0.001 |
| %Bachelor's or higher | 0.192 | 0.188 | 0.004 | 0.055 |
| Violent Crime Rate | 503.651 | 506.470 | -2.819 | 0.587 |
| Log of Coincident | 4.534 | 4.561 | -0.027 | 0.000 |
| Incumbent Runs is in the Contest | 0.486 | 0.533 | -0.047 | 0.000 |
| Observations | 10918 | | | |

The 'Control' group consists of contests with no disaster 0-12 months prior. The 'Treated' group is those with a disaster. 'Difference' is Treated - Control. P-value is from an independent t-test.

**Table 3: Joint Significance Test: Non-Incumbent Pre-Treatment Coefficients**

| Hypothesis | F(3, 47) | p-value |
|---|---|---|
| 1.relq0 = 0, 2.relq0 = 0, 3.relq0 = 0 | 1.32 | 0.2790 |

*Interpretation: Fail to reject null: No significant pre-existing trends in non-incumbent group.*

**Table 4: Joint Significance Test: Interaction Terms (Incumbent Differential Pre-Trends)**

| Hypothesis | $F(3, 47)$ | p-value |
|---|---|---|
| 1.relq0#1.incumbent = 0, 2.relq0#1.incumbent = 0, 3.relq0#1.incumbent = 0 | 0.82 | 0.4894 |

*Interpretation: Fail to reject null: No differential pre-trend between incumbents and non-incumbents.*

# Table 5: Effect of Pre-Election Disaster on Vote Share and Winning the election

Effect of Pre-Election Disaster on Vote Share and Winning the election

|  | (1) Vote share | (2) Vote share | (3) Wins the election | (4) Wins the election |
|---|---|---|---|---|
| 1.treated | 0.001 (0.08) | -0.015 (-0.86) | 0.001 (0.10) | -0.017 (-1.00) |
| incumbent=1 | 0.296*** (18.23) | 0.283*** (8.32) | 0.511*** (19.63) | 0.478*** (12.73) |
| 1.treated # incumbent=1 | 0.019 (1.17) | 0.032 (0.96) | 0.022 (0.99) | 0.058 (1.63) |
| democrat | 0.082*** (8.31) | 0.081*** (7.50) | 0.130*** (6.62) | 0.125*** (5.75) |
| male | 0.014** (2.27) | 0.013** (2.61) | 0.030** (2.25) | 0.028** (2.49) |
| white | 0.034*** (4.01) | 0.031*** (3.39) | 0.050*** (3.52) | 0.048*** (3.12) |
| male% | 0.067 (0.12) | 0.120 (0.20) | -0.011 (-0.02) | 0.011 (0.02) |
| white% | 0.098 (1.65) | 0.080 (1.56) | 0.151** (2.33) | 0.142** (2.40) |
| % In labor force | 0.133 (1.03) | 0.145 (0.93) | 0.153 (1.23) | 0.184 (1.19) |
| %Bachelor's or higher | -0.148 (-1.11) | -0.122 (-1.14) | -0.089 (-0.60) | -0.078 (-0.56) |
| Violent Crime Rate | 0.000 (1.18) | 0.000 (1.28) | 0.000 (1.33) | 0.000 (1.63) |
| Log of Coincident | -0.062 (-1.32) | -0.060 (-1.36) | -0.051 (-1.09) | -0.044 (-0.93) |
| Log of Population | 0.012 | 0.019 | -0.015 | -0.010 |
| | (0.58) | (1.03) | (-0.90) | (-0.71) |
| Constant | 0.242 | 0.139 | 0.405 | 0.297 |
| | (0.70) | (0.40) | (1.23) | (0.97) |
| Observations | 10593 | 9106 | 10593 | 9106 |

*t-statistics* in parentheses

Odd columns' Control group includes post-election and no-disaster observations, and even columns' Control group includes only no-disaster observations. Year and FIPS Fixed Effects. SEs clustered by state.

$^{*} p < 0.10$, $^{**} p < 0.05$, $^{***} p < 0.01$

**Table 6: Heterogeneous Effects on Vote Share and Winning the election**

Heterogeneous Effects on Vote Share and Winning the election

|  |  | (1) Vote share | (2) Vote share | (3) Wins the election | (4) Wins the election |
|---|---|---|---|---|---|
| 1.quarters_before |  | -0.007 (-0.40) | -0.022 (-1.06) | -0.011 (-0.55) | -0.028 (-1.28) |
| 2.quarters_before |  | 0.012 (0.99) | -0.005 (-0.29) | 0.011 (0.81) | -0.008 (-0.44) |
| 3.quarters_before |  | -0.011 (-0.86) | -0.027 (-1.61) | -0.012 (-0.85) | -0.031* (-1.74) |
| 4.quarters_before |  | 0.007 (0.45) | -0.009 (-0.46) | 0.009 (0.64) | -0.007 (-0.42) |
| incumbent=1 |  | 0.296*** (18.21) | 0.283*** (8.34) | 0.511*** (19.62) | 0.479*** (12.76) |
| 1.quarters_before incumbent=1 | # | 0.052** (2.44) | 0.064* (1.80) | 0.074** (2.63) | 0.110** (2.66) |
| 2.quarters_before incumbent=1 | # | 0.022 (1.06) | 0.037 (1.02) | 0.050 (1.42) | 0.090* (1.93) |
| 3.quarters_before incumbent=1 | # | 0.018 (1.08) | 0.029 (0.86) | 0.008 (0.28) | 0.041 (0.97) |
| 4.quarters_before incumbent=1 | # | 0.007 (0.36) | 0.020 (0.56) | 0.004 (0.16) | 0.039 (1.11) |
| democrat |  | 0.082*** (8.32) | 0.081*** (7.50) | 0.130*** (6.63) | 0.125*** (5.76) |
| male |  | 0.014** (2.27) | 0.014** (2.62) | 0.030** (2.27) | 0.029** (2.53) |
| white |  | 0.034*** | 0.031*** | 0.050*** | 0.048*** |

|  | (4.04) | (3.41) | (3.53) | (3.12) |
|---|---|---|---|---|
| male% | 0.066 | 0.116 | -0.023 | -0.008 |
|  | (0.12) | (0.19) | (-0.04) | (-0.01) |
| white% | 0.101* | 0.084 | 0.154** | 0.146** |
|  | (1.72) | (1.64) | (2.41) | (2.49) |
| % In labor force | 0.147 | 0.163 | 0.173 | 0.212 |
|  | (1.16) | (1.06) | (1.43) | (1.39) |
| %Bachelor's or higher | -0.140 | -0.113 | -0.083 | -0.069 |
|  | (-1.05) | (-1.05) | (-0.55) | (-0.50) |
| Violent Crime Rate | 0.000 | 0.000 | 0.000 | 0.000* |
|  | (1.33) | (1.41) | (1.48) | (1.76) |
| Log of Coincident | -0.065 | -0.063 | -0.055 | -0.048 |
|  | (-1.42) | (-1.46) | (-1.20) | (-1.04) |
| Log of Population | 0.012 | 0.019 | -0.016 | -0.010 |
|  | (0.61) | (1.08) | (-0.95) | (-0.78) |
| Constant | 0.241 | 0.136 | 0.416 | 0.307 |
|  | (0.71) | (0.40) | (1.29) | (1.03) |
| Observations | 10593 | 9106 | 10593 | 9106 |

*t-statistics* in parentheses

Odd columns' Control group includes post-election and no-disaster observations, and even columns' Control group includes only no-disaster observations. Year and FIPS Fixed Effects. SEs clustered by state.

* $p < 0.10$, ** $p < 0.05$, *** $p < 0.01$

**Table 7: Effect of Pre-Election Disaster on Incumbent Margin**

Effect of Pre-Election Disaster on Incumbent Margin

|  | (1) Margin | (2) Margin |
|---|---|---|
| 1.treated | 0.021 (1.22) | 0.063* (1.84) |
| democrat | 0.028 (0.75) | 0.036 (0.78) |
| male | 0.052 (1.66) | 0.046 (1.41) |
| white | -0.002 (-0.07) | 0.008 (0.25) |
| male% | 0.527 (0.38) | 0.836 (0.50) |
| white% | 0.011 (0.08) | 0.001 (0.01) |
| % In labor force | -0.023 (-0.06) | -0.222 (-0.47) |
| %Bachelor's or higher | -0.153 (-0.28) | 0.147 (0.23) |
| Violent Crime Rate | -0.000 (-0.95) | -0.000 (-1.12) |
| Log of Coincident | -0.207 (-0.80) | -0.168 (-0.58) |
| Log of Population | 0.040 (0.77) | 0.017 (0.30) |
| Constant | 0.495 (0.41) | 0.455 (0.31) |
| Observations | 1718 | 1460 |

*t-statistics* in parentheses

Dep. Var: margin. Odd columns' Control group includes post-election and no-disaster observations, and even columns' Control group includes only no-disaster observations. Year and

FIPS Fixed Effects. SEs clustered by state.
$^{*} p < 0.10, ^{**} p < 0.05, ^{***} p < 0.01$

## Table 8: Heterogeneous Effects on Incumbent Margin

Heterogeneous Effects on Incumbent Margin

|  | (1) Margin | (2) Margin |
|---|---|---|
| 1.quarters_before | 0.052** (2.09) | 0.095** (2.49) |
| 2.quarters_before | 0.015 (0.53) | 0.065 (1.52) |
| 3.quarters_before | 0.032 (1.41) | 0.073* (1.99) |
| 4.quarters_before | -0.011 (-0.55) | 0.030 (0.79) |
| democrat | 0.029 (0.76) | 0.036 (0.79) |
| male | 0.054* (1.70) | 0.049 (1.47) |
| white | -0.004 (-0.13) | 0.005 (0.16) |
| male% | 0.520 (0.37) | 0.849 (0.51) |
| white% | 0.007 (0.05) | -0.008 (-0.06) |
| % In labor force | -0.034 (-0.09) | -0.247 (-0.54) |
| %Bachelor's or higher | -0.178 (-0.32) | 0.111 (0.18) |
| Violent Crime Rate | -0.000 (-1.17) | -0.000 (-1.36) |
| Log of Coincident | -0.200 (-0.79) | -0.163 (-0.57) |
| Log of Population | 0.035 (0.66) | 0.008 (0.14) |
| Constant | 0.558 | 0.577 |
| | (0.46) | (0.40) |
| Observations | 1718 | 1460 |

*t-statistics* in parentheses

Dep. Var: margin. Odd columns' Control group includes post-election and no-disaster observations, and even columns' Control group includes only no-disaster observations. Year and FIPS Fixed Effects. SEs clustered by state.

$^*$ $p < 0.10$, $^{**}$ $p < 0.05$, $^{***}$ $p < 0.01$

## Table 9: Event Study Results

Event Study Results

| | (1) Vote Share |
|---|---|
| 1.relq0 | 0.057 |
| | (0.035) |
| 2.relq0 | -0.026 |
| | (0.029) |
| 3.relq0 | 0.007 |
| | (0.021) |
| 5.relq0 | -0.006 |
| | (0.017) |
| 6.relq0 | 0.018 |
| | (0.013) |
| 7.relq0 | -0.005 |
| | (0.012) |
| 8.relq0 | 0.015 |
| | (0.016) |
| 1.incumbent | 0.288*** |
| | (0.015) |
| 1.relq0#1.incumbent | -0.007 |
| | (0.046) |
| 2.relq0#1.incumbent | 0.046 |
| | (0.060) |
| 3.relq0#1.incumbent | 0.050 |
| | (0.038) |
| 5.relq0#1.incumbent | 0.063*** |
| | (0.020) |
| 6.relq0#1.incumbent | 0.027 |
| | (0.020) |
| 7.relq0#1.incumbent | 0.026 |
|  | (0.016) |
| 8.relq0#1.incumbent | 0.014 |
|  | (0.019) |
| democrat | 0.084*** |
|  | (0.010) |
| male | 0.014** |
|  | (0.007) |
| white | 0.034*** |
|  | (0.009) |
| Violent Crime Rate | 0.000 |
|  | (0.000) |
|  | (0.043) |
| male% | -0.104 |
|  | (0.572) |
| white% | 0.068 |
|  | (0.063) |
| % In labor force | 0.216 |
|  | (0.130) |
| %Bachelor's or higher | -0.169 |
|  | (0.130) |
| Log of Population | 0.008 |
|  | (0.022) |
| _cons | 0.508 |
|  | (0.307) |
| $N$ | 9755 |

Standard errors in parentheses

Notes: Standard errors clustered by state. Year and FIPS fixed effects included.

$^{*}\ p < 0.10$, $^{**}\ p < 0.05$, $^{***}\ p < 0.01$

**Table 10: Effect of Pre-Election Disaster on Voter Turnout and Number of Candidates**

Effect of Pre-Election Disaster on Voter Turnout and Number of Candidates

|  | (1) Voter Turnout Rate | (1) Number of Candidates in Contest |
|---|---|---|
| 1.treated | -0.014** (-2.15) | -0.134 (-0.82) |
| Incumbent Runs | -0.006 (-1.40) | -0.387** (-2.48) |
| 1.treated # 1. Incumbent Runs | 0.007 (0.99) | 0.200 (1.18) |
| Share of White Candidates in Contest | -0.010* (-1.77) | -0.286*** (-3.18) |
| Share of Male Candidates in Contest | -0.003 (-0.66) | -0.145* (-1.95) |
| male% | -0.558*** (-3.30) | -3.980 (-1.01) |
| white% | 0.034 (1.30) | -0.212 (-0.42) |
| % In labor force | -0.108 (-1.16) | 0.907 (0.78) |
| %Bachelor's or higher | 0.324*** (4.75) | 2.605 (1.52) |
| Violent Crime Rate | 0.000 (1.53) | -0.000 (-1.22) |
| Log of Coincident | 0.144*** (3.36) | 0.629* (1.73) |
| Log of Population | -0.058*** (-3.85) | -0.544*** (-3.26) |
| Constant | 0.440* (1.98) | 7.890*** (2.86) |
| Observations | 4279 | 4279 |

*t-statistics* in parentheses

The control group includes post-election and no-disaster contests. Year and FIPS Fixed Effects. SEs clustered by state.

* $p < 0.10$, ** $p < 0.05$, *** $p < 0.01$

**Table 11: Heterogeneous Effects on Voter Turnout and Number of Candidates by Disaster Timing**

| Heterogeneous Effects on Voter Turnout and Number of Candidates by Disaster Timing | | |
|---|---|---|
| | (1) Voter Turnout Rate | (2) Number of Candidates in Contest |
| 1.quarters_before | -0.020* <br> (-1.72) | 0.009 <br> (0.04) |
| 2.quarters_before | -0.014 <br> (-1.50) | -0.145 <br> (-0.81) |
| 3.quarters_before | -0.009 <br> (-1.66) | 0.052 <br> (0.26) |
| 4.quarters_before | -0.013** <br> (-2.52) | -0.283 <br> (-1.34) |
| Incumbent Runs | -0.006 <br> (-1.46) | -0.380** <br> (-2.42) |
| 1.quarters_before # 1. Incumbent Runs | 0.003 <br> (0.26) | 0.071 <br> (0.29) |
| 2.quarters_before # 1. Incumbent Runs | 0.005 <br> (0.61) | 0.037 <br> (0.21) |
| 3.quarters_before # 1. Incumbent Runs | 0.005 <br> (1.19) | -0.001 <br> (-0.00) |
| 4.quarters_before # 1. Incumbent Runs | 0.009 <br> (1.10) | 0.367* <br> (1.86) |
| Share of White Candidates in Contest | -0.010* <br> (-1.69) | -0.290*** <br> (-3.24) |
| Share of Male Candidates in Contest | -0.003 <br> (-0.79) | -0.142* <br> (-1.85) |
| male% | -0.546*** <br> (-3.31) | -4.077 <br> (-1.02) |
| white% | 0.034 <br> (1.30) | -0.219 <br> (-0.43) |
| % In labor force | -0.117 | 0.804 |
| | (-1.23) | (0.71) |
| %Bachelor's or higher | 0.327*** | 2.611 |
| | (4.82) | (1.55) |
| Violent Crime Rate | 0.000 | -0.000 |
| | (1.52) | (-1.34) |
| Log of Coincident | 0.141*** | 0.668* |
| | (3.35) | (1.73) |
| Log of Population | -0.057*** | -0.562*** |
| | (-3.81) | (-3.29) |
| Constant | 0.435* | 8.047*** |
| | (2.00) | (2.89) |
| Observations | 4279 | 4279 |

*t-statistics* in parentheses
The base group is post-election and no-disaster contests. Year and FIPS Fixed Effects. SEs clustered by state.
* $p < 0.10$, ** $p < 0.05$, *** $p < 0.01$

# Table 12: Effect of Inter-Election Disasters on Candidate's Future Participation

| Effect of Inter-Election Disasters on Candidate's Future Participation | |
|---|---|
| | (1) |
| | Will the Candidate Participate in the next election? |
| 1.years_after | 0.026 |
| | (0.73) |
| 2.years_after | 0.060 |
| | (1.51) |
| 3.years_after | 0.013 |
| | (0.25) |
| incumbent=1 | -0.012 |
| | (-0.25) |
| 1.years_after # incumbent=1 | 0.121** |
| | (2.38) |
| 2.years_after # incumbent=1 | 0.104* |
| | (1.90) |
| 3.years_after # incumbent=1 | -0.011 |
| | (-0.12) |
| democrat | 0.113*** |
| | (8.78) |
| male | 0.026*** |
| | (2.85) |
| white | 0.005 |
| | (0.43) |
| male% | 0.736 |
| | (1.16) |
| white% | -0.113* |
| | (-1.81) |
| %In labor force | -0.091 |
|   | (-0.37) |
| %Bachelor's or higher | -0.122 |
|   | (-0.70) |
| Violent Crime Rate | -0.000 |
|   | (-1.04) |
| Log of Coincident | 0.008 |
|   | (0.12) |
| Log of Population | 0.036 |
|   | (1.46) |
| Constant | -0.464 |
|   | (-0.99) |
| Observations | 10593 |

$t$ statistics in parentheses
Dep. Var: will_participate_next. Base group is no disaster in 3 years post-election. Year and FIPS Fixed Effects. SEs clustered by state.
$^{*} p < 0.10$, $^{**} p < 0.05$, $^{***} p < 0.01$

# Table 13: Placebo Test using Post-Election Disasters on Vote Share and Margin

| Placebo Test using Post-Election Disasters on Vote Share and Margin | | |
|---|---|---|
| | (1) Vote share | (2) Margin |
| 1.quarters_after | 0.024 (1.14) | 0.078 (1.49) |
| 2.quarters_after | -0.000 (-0.01) | 0.145 (1.12) |
| 3.quarters_after | -0.003 (-0.16) | 0.127 (1.38) |
| 4.quarters_after | 0.053 (1.12) | 0.264 (1.49) |
| incumbent=1 | 0.289*** (7.76) | |
| 1.quarters_after # incumbent=1 | -0.001 (-0.01) | |
| 2.quarters_after # incumbent=1 | 0.019 (0.29) | |
| 3.quarters_after # incumbent=1 | 0.020 (0.41) | |
| 4.quarters_after # incumbent=1 | -0.005 (-0.07) | |
| democrat | 0.072*** (6.43) | -0.037 (-0.71) |
| male | 0.008 (0.61) | 0.034 (0.59) |
| white | 0.039*** (3.35) | -0.011 (-0.14) |
| male% | -0.654 | 4.005 |
|  | (-0.67) | (1.61) |
| white% | 0.304* | -0.332 |
|  | (1.81) | (-0.68) |
| % In labor force | 0.128 | 1.563 |
|  | (0.41) | (1.08) |
| %Bachelor's or higher | 0.271 | -2.950** |
|  | (0.90) | (-2.22) |
| Violent Crime Rate | -0.000 | -0.000 |
|  | (-0.31) | (-0.14) |
| Log of Coincident | -0.014 | -0.328 |
|  | (-0.13) | (-0.99) |
| Log of Population | -0.023 | 0.023 |
|  | (-0.30) | (0.17) |
| Constant | 0.598 | -0.565 |
|  | (0.60) | (-0.28) |
| Observations | 2295 | 244 |

*t-statistics* in parentheses
The base group is no-disaster obs only. Year and FIPS Fixed Effects. SEs clustered by state.
* $p < 0.10$, ** $p < 0.05$, *** $p < 0.01$